\documentstyle[aps,prb,amssymb,epsfig,times]{revtex}

\begin{document}

\twocolumn[%
 \csname@twocolumnfalse\endcsname
 \hsize\textwidth\columnwidth\hsize

\title{Spatiotemporal Stochastic Resonance in Fully Frustrated Josephson
Ladders}
\author{Beom Jun Kim$^a$, Mahn-Soo Choi$^{b,c}$, Petter Minnhagen$^a$,
    Gun Sang Jeon$^{c}$~\cite{gun}, H.J. Kim$^{d}$, and M.Y. Choi$^{d}$}
\address{$^a$Department of Theoretical Physics, Ume\aa\ University, 901 87
    Ume\aa, Sweden\\
    $^b$Departement Physik und Astronomie, Universit\"at Basel,
    Klingelbergstrasse 82, CH-4056 Basel, Switzerland \\
    $^c$School of Physics, Korea Institute for Advanced Study,
    Cheongryanri-dong 207-43, Seoul 130-012, Korea \\
    $^d$Department of Physics and Center for Theoretical Physics, Seoul
    National University, Seoul 151-742, Korea}
\maketitle
\draft

\begin{abstract}
We consider a Josephson-junction ladder in an external magnetic field with
half flux quantum per plaquette. When driven by external currents, periodic
in time and staggered in space, such a fully frustrated system is found to
display spatiotemporal stochastic resonance under the influence of thermal
noise.  Such resonance behavior is investigated both numerically and
analytically, which reveals significant effects of anisotropy and yields
rich physics.
\end{abstract}


\bigskip

]%


\section{Introduction}

In a nonlinear system driven by a weak periodic signal, noise present in
the system can operate cooperatively with the input signal.  This effect,
known as {\em stochastic resonance} (SR), emerges when the {\em stochastic}
time scale of the noise matches the {\em deterministic} time scale of the
input signal.  Since its introduction two decades ago or so,~\cite{Benzi}
SR has been observed in a wide range of systems.~\cite{Gammaitoni} Recent
interest has been attracted to the collective spatial and temporal behavior
of SR in arrays of locally or globally coupled nonlinear oscillators and in
spatially extended continuous systems under the influence of noise and weak
periodic driving.~\cite{Lindner,Locher} In particular, it has been shown
that the SR in an array of locally and linearly coupled bistable overdamped
oscillators can significantly be enhanced with respect to that in  a single
{\em stochastic resonator}.~\cite{Lindner} The optimal coupling and noise
strength scale with the system size, and the signal-to-noise ratio (SNR) in
the optimal conditions is improved by a considerable amount.  Such effects
of array enhanced stochastic resonance (AESR) have also been observed
experimentally in a system of coupled diode resonators.~\cite{Locher} The
AESR and other interesting aspects of the spatiotemporal stochastic
resonance (STSR) are believed to have potential importance in the area of
signal and image processing and pattern formation.

Most of the works on STSR have been focused on model systems, which are
relatively easy to study numerically and sometimes analytically (cf.
Ref.~\onlinecite{Locher}).  For instance, widely studied are the
one-dimensional time-dependent Ginzburg-Landau model~\cite{Benzi2} and the
Langevin-type stochastic equations for linearly coupled bistable
oscillators.~\cite{Lindner} On one hand, they are rather general in nature
and applicable to various systems; on the other hand, relevant parameters
can only be determined phenomenologically for a specific system.  Another
class of model worth mentioning is the kinetic Ising model.~\cite{book}
Although the Ising model {\it per se} does not have any a priori dynamics, it can
be provided with an artificially defined temporal evolution consistent
with the equilibrium properties of the Ising magnets.  While
the kinetic Ising model has been extensively studied mainly via Monte Carlo
simulation methods in various contexts including STSR,~\cite{Neda} it is
not clear a priori whether the kinetic Ising model describes the true
dynamics of a real physical system, especially, far from equilibrium.

\begin{figure}
\centerline{\epsfig{file=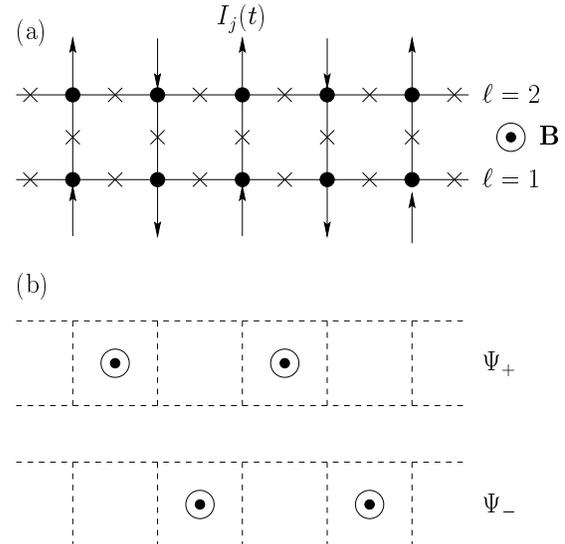,width=8cm}}
\caption{(a) Schematic diagram of the fully frustrated Josephson-junction
ladder (FFJJL) driven by staggered currents.
(b) Two degenerate ground states of FFJJL in the absence of external
current, with $\odot$'s indicating the positions of vortices.
}
\label{fig:1}
\end{figure}

In this paper, we consider a Josephson-junction ladder (JJL), two
one-dimensional arrays of superconducting grains coupled with each other
via Josephson junctions, in an external transverse magnetic field with half
flux quantum per plaquette (see Fig.~\ref{fig:1}).  It has two degenerate
ground states and switches back and forth from one ground state to the
other under the influence of thermal noise and the periodic (in time) and
{\em staggered} (in space) external current driving.  Such a fully
frustrated Josephson-junction ladder (FFJJL) has several merits as an
example of the real physical system to investigate STSR in: The temporal
evolution of the system follows the well-known resistively shunted junction
(RSJ) dynamics,~\cite{Tinkham} described by a set of overdamped coupled
stochastic differential equations for phase variables on superconducting
grains.  Each parameter in the RSJ model has a clear physical meaning and can
be controlled easily in experiment.~\cite{Benz} Moreover, the JJL itself
has been a topic of a great number of studies due to its rich physics in
conjunction with vortex motion and frustration
effects.~\cite{Benz,Granato,Denniston,Kim} Here we perform numerical
simulations on the RSJ model, and examine the spatiotemporal behavior of SR
in the FFJJL.  Of particular interest are the effects on STSR of the
anisotropy in the Josephson coupling along the leg- and the rung-direction,
which turns out to yield quite rich physics.  We also present an analytical
investigation, based on the Glauber dynamics of vortices, and make
comparison with the simulation results. 

This paper is organized as follows: In Sec. II we introduce the model and
specify associated control parameters in the system.  The model is also
related to a one-dimensional system of vortices and to the
antiferromagnetic Ising model in a time-dependent staggered magnetic field.
A part of the derivation of these connections is contained in Appendix A.
Section III presents results of numerical simulations.  We first
characterize the SR behavior of the system, and investigate the anisotropy
effects on the STSR behavior.  These simulation results are compared with
the analytical ones in Sec. IV, obtained from the Glauber dynamics of
vortices.  Finally, Section V concludes the paper with a brief discussion
about possible experimental setups.

\section{Model}

Within the RSJ model (neglecting the capacitive and inductive effects), a
Josephson junction can be characterized by the Josephson critical current
$I_J$ or equivalently the Josephson coupling $J\equiv \hbar I_J / 2e$ and
the shunt resistance $R$, which define the frequency scale $\omega_J \equiv
2e R I_J / \hbar$ for the dynamics of the system.  Throughout this paper,
we always measure energy and temperature in units of $J$, time and
frequency with respect to $\omega_J$, and other induced quantities
correspondingly.  A JJL of length $L$ in a transverse external magnetic
field, depicted in Fig.~\ref{fig:1}, is then described by the set of
coupled stochastic nonlinear differential equations for phase variables
\begin{equation} \label{motion}
\sum_{i\in N(j)} \left[ \dot{\phi}_{ij} (t) + J_{ij} \sin (\phi_{ij} -
A_{ij} ) + \zeta_{ij} (t) \right] = I_j (t),
\end{equation}
where $\phi_{ij} \equiv \phi_i - \phi_j $ is the phase difference across
the junction $(ij)$, $N(j)$ denotes the set of nearest neighbors of site
$j$, and $\zeta_{ij} (t)$ is the random noise current on the junction with
zero average and correlation
\begin{equation} \label{noise}
\langle \zeta_{ij} (t) \zeta_{i'j'} (t') \rangle 
= 2 T [\delta_{(ij),(i'j')} - \delta_{(ij),(j'i')} ] \delta(t-t').
\end{equation}
The Josephson coupling strength $J_{ij} \,(= J_{ji})$ is given by
\begin{equation}
J_{ij} = \left\{ 
\begin{array}{cl}
\alpha & \hbox{for }(ij) \hbox{ legs},\\
1 & \hbox{for }(ij) \hbox{ rungs},
\end{array}
\right.
\end{equation}
where we have introduced the anisotropy factor $\alpha$ between leg-junctions 
(abbreviated as legs) and rung-junctions (rungs).~\cite{endnote}
The external current in Eq.~(\ref{motion}) is periodic in time with frequency
$\Omega$ and staggered in space
\begin{equation} \label{current}
I_j (t) = (-1)^{\ell + x} I_0 \cos \Omega t ,
\end{equation}
where each site index $j$ has been further specified by the leg index $\ell
\,(= 1, 2)$ and the position $x$ along the leg, i.e., $j\equiv(\ell,x)$.
The effects of the magnetic field are manifested in the gauge field $A_{ij}
= - A_{ji}$, which, in the Landau gauge, takes the form
\begin{equation}
A_{ij} = \left\{ 
\begin{array}{cl}
0 & \hbox{for }(ij) \hbox{ legs},\\
2\pi f x & \hbox{for }(ij) \hbox{ rungs},
\end{array}
\right. 
\end{equation}
where $f \equiv \Phi / \Phi_0$ is the flux $\Phi$ per plaquette in
units of the flux quantum $\Phi_0 \equiv hc / 2e$.

Most of the physical properties of a Josephson junction ladders (and arrays
in general) can be understood most clearly in terms of vortices.
Accordingly, we introduce an effective vortex Hamiltonian,
\begin{equation} \label{VHam2}
H_V
= \sum_{x,x'} (v_x - f - \delta f_x)G(x-x')(v_{x'}-f-\delta f_{x'}), 
\end{equation}
which is equivalent to Eq.~(\ref{motion})
within the Villain scheme (see Appendix A for details).
In Eq.~(\ref{VHam2}), the {\em vorticity} $v_x$ on plaquette $x$ is given
by the directional sum of the gauge-invariant phase differences around the
plaquette and only takes the values $\pm 1$
while the effective interaction $G(x)$ between vortices obtains
the form 
\begin{equation} \label{Gx}
G(x)
= \frac{\pi^2 \alpha}{\sqrt{1+2\alpha}}
  \left[\frac{1 + \alpha + \sqrt{1+2\alpha}}{\alpha}\right]^{-|x|}
\end{equation}
in the limit of $L\rightarrow\infty$.
The additional time-dependent flux $\delta f_x(t)$ is due to the external
driving current and given by
\begin{equation} \label{tdep}
\delta f_x(t)
= \frac{2 \alpha -1}{\pi(1+4\alpha^2)} (-1)^x I_0 \cos\Omega t \;.
\end{equation}

For the purpose of discussions below, one important point here is the
following:  Without external current driving ($I_0=\delta f_x=0$) and at the
maximal frustration ($f=1/2$), the Hamiltonian (\ref{VHam2}) has two
degenerate ground state $\Phi_\pm$ [see Fig.1(b)]
with different values of the staggered magnetization
\begin{equation}\label{staggeredmag}
m \equiv \frac{1}{L} \sum_x (-1)^x \langle v_x-\frac{1}{2} \rangle
= \pm\frac{1}{2} \;.
\end{equation}
The additional flux favors one of the two states $\Psi_\pm$ and thus lifts
the degeneracy.  Thus the system under consideration is somewhat
analogous to a simple two state model and from this perspective
the existence of a SR response may be intuitively expected.
It should be noted here that the sign of $\delta f_x$
depends on the anisotropy factor $\alpha$, implying that the choice between
the two states $\Psi_\pm$ depends on the value of $\alpha$: For
$\alpha<\alpha_{cr}^0\equiv{1/2}$ the system favors, say, the state
$\Psi_+$ and for $\alpha>\alpha_{cr}^0$ the other state $\Psi_-$.  This can
be understood from the fact that the external currents tend to flow through
the easier links, i.e., the rungs for small $\alpha$ and the legs for large
$\alpha$.

In the following sections, we will use the more complete and accurate model
(\ref{motion}) for numerical simulations, whereas the effective model
(\ref{VHam2}) will be the starting point of our analytic approach.

\section{Simulation Results}

We first present results from the simulations of the RSJ dynamics.
The set of equations of motion in Eq.~(\ref{motion}) is integrated 
through the use of the simple Euler algorithm 
with discrete time steps $\Delta t = 0.05$, 
and the thermal noise currents satisfying Eq.~(\ref{noise}) are generated 
to follow the uniform distribution.
To use the efficient tridiagonal matrix algorithm,~\cite{Press}
we separate Eq.~(\ref{motion}) into two parts by change of variables (see
Ref.~\onlinecite{Kim} for details).
For the staggered input current, we set $I_0 =0.1$ and $\Omega=0.002$.

\begin{figure}
\centerline{\epsfig{file=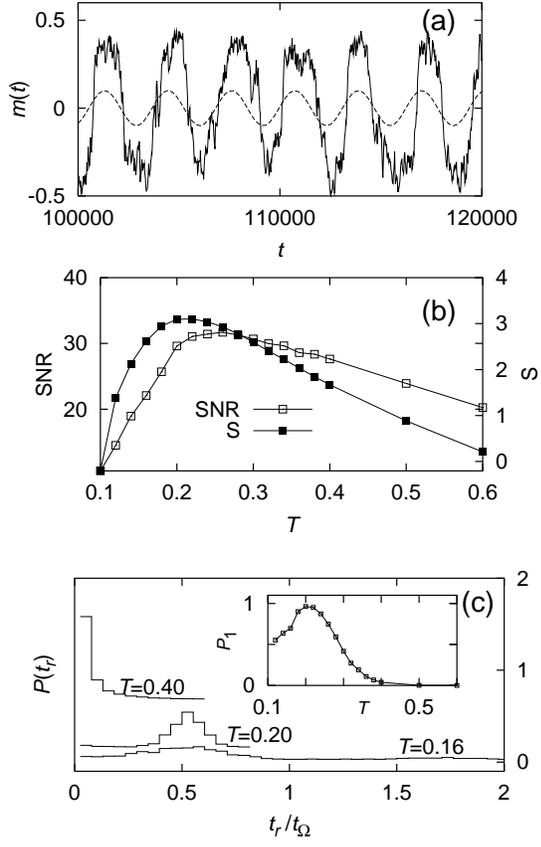,width=8cm}}
\caption{Various characteristics of stochastic resonance in the system of
size $L=128$ for the driving frequency $\Omega=0.002$:
(a) time-series of the staggered magnetization $m(t)$ (solid line) and the
external current (dashed line) at $T=0.26$; 
(b) signal-to-noise ratio SNR (left vertical scale) and output signal power
S at frequency $\Omega$ (right scale) as functions of the temperature;
(c) residence-time distribution at several temperatures [inset: primary
peak height in the residence-time distribution as a function of the
temperature].}
\label{fig:2}
\end{figure}

In Fig.~\ref{fig:2}, we characterize typical SR behaviors of 
the isotropic system ($\alpha =1$)
with size $L=128$ in three different ways: 
(a) the time series of the staggered magnetization, 
(b) the signal-to-noise ratio (SNR), and (c) the residence-time distribution. 
In Fig.~\ref{fig:2}(a), the staggered magnetization $m(t) \equiv
(1/L)\sum_x (-1)^x m_x(t)$ [equivalent to Eq.~(\ref{staggeredmag})], 
where $m_x$ has been obtained from the plaquette sum of 
the gauge-invariant phase differences,
shows clear synchronization to the input driving
current at $T=0.26$.
Such behavior is observed only in the intermediate range of temperatures.
At sufficiently low temperatures the system does not
follow the external driving but most of the time stays in one of its two stable
states ($m(t) = \pm 1/2$), while at sufficiently high temperatures the
resonance effect is washed out by strong thermal fluctuations.
The SR behavior is also clearly reflected in the power spectrum
defined by 
\begin{equation}
S(\omega) \equiv \left\langle \left| \int_0^\Theta dt e^{-i\omega t} m(t) m(0)
\right|^2 \right\rangle ,
\end{equation}
where $\langle \cdots \rangle$ denotes the average over thermal noises, 
and the values $m(0)= -1/2$ and $\Theta = 50 t_\Omega$ 
with $t_\Omega \equiv 2\pi/ \Omega $ have been used in the simulations.
The SNR is then defined by
\begin{equation} \label{SNR}
\hbox{SNR} \equiv 10 \log_{10} \left[ {S \over N(\Omega)} \right],
\end{equation}
where $S$ is the output signal strength given by the area under the peak at
$\omega = \Omega$ and $N(\Omega)$ is the noise level at $\omega=\Omega$.
The SNR and the signal strength $S$ [Fig.~\ref{fig:2}(b)] display peaks at
$T\simeq 0.26$ and $T\simeq 0.22$, respectively.
It is clear from the time series of the staggered magnetization $m(t)$
[Fig.~\ref{fig:2}(a)] and also the residence-time distribution
[Fig.~\ref{fig:2}(c), see below] that these peaks are associated with the
global synchronization of the system to the external periodic driving.
As an alternative characterization of the SR behavior, we also measure the
residence time $t_r$, i.e., how long the system stays in one of its
stable states,~\cite{Gammaitoni} and show in
Fig.~\ref{fig:2}(c) its distribution $P(t_r )$ at various temperatures.
Near $T=0.2$, the distribution develops a narrow peak 
around $t_r = t_\Omega/2$, again signaling the input/output synchronization.
The primary peak height $P_1$,
defined by the area of the distribution around $t_r = t_\Omega/2$,
is shown in the inset of Fig.~\ref{fig:2}(c), manifesting the
broad peak near $T\approx 0.2$.

\begin{figure}
\centerline{\epsfig{file=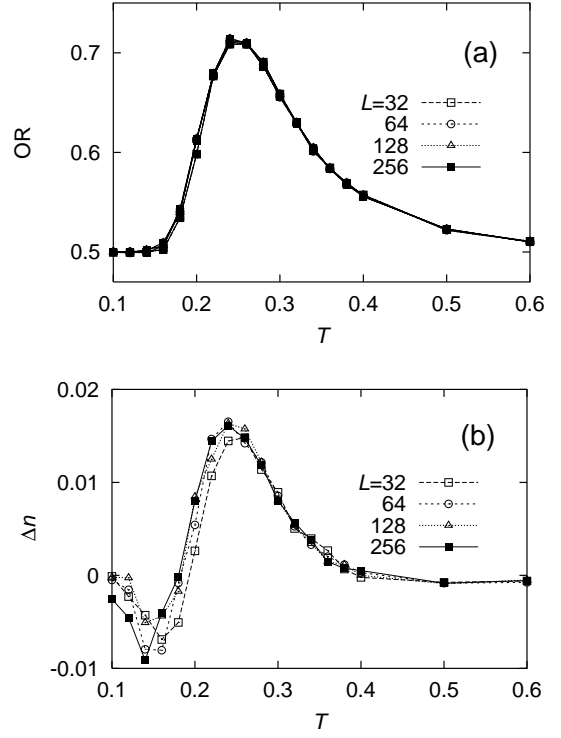,width=8cm}}
\caption{(a) Occupancy ratio OR and (b) deviation in the average number of
domain walls from the equilibrium value, $\Delta n \equiv n(0) - n(I_0)$,
for $\alpha=1$ and $L=$32, 64, 128, and 256.}
\label{fig:3}
\end{figure}

Near SR, synchronization of the magnetization on each plaquette leads to
the enhancement of the spatial correlation in the system, 
as demonstrated in Fig.~\ref{fig:3}.
In Fig.~\ref{fig:3}(a), we plot the occupancy ratio (OR),~\cite{Lindner}
which measures how many plaquettes in the
ladder follow (in phase) the external driving.
In our case it is given by
\begin{equation} \label{OR}
\hbox{OR} = {1\over2} \langle\!\langle 1 + m (t_+ ) - m(t_- ) \rangle\!\rangle,
\end{equation}
where $\cos (\Omega t_{\mp} ) {{}_{\hbox{$<$}} \atop {}^{\hbox{$>$}}} 0$ 
and $\langle\!\langle \cdots \rangle\!\rangle$ represents 
the averages over time and ensemble. 
Note that the temperature at which the maximum occupancy is attained 
coincides with the temperature where the SNR develops a peak.
To show the enhancement of the spatial correlation in a more direct way,
we plot in Fig.~\ref{fig:3}(b) the deviation of the average number of
domain walls for staggered magnetization from the equilibrium value, 
$\Delta n(I_0) \equiv n(0) -n(I_0)$.
Reduction in the number of domain walls near the resonance temperature $T_{SR}$
can be observed. 
Here the initial suppression of the spatial correlation (for $T \lesssim 0.2$) 
is due to the asynchronization of the system: 
With each plaquette out of phase by a random amount, 
the effect of the external driving is similar to that of random disorder.

\begin{figure}
\centerline{\epsfig{file=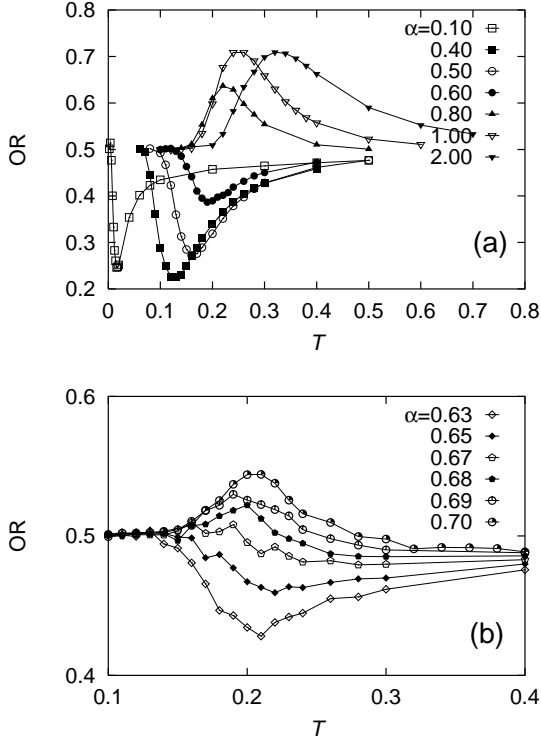,width=8cm}}
\caption{Occupancy ratio for different values of the anisotropy factor
$\alpha$.  The lower panel (b) focuses on the region $\alpha \simeq
\alpha_{cr}$ and $T \simeq T_{SR}$.}
\label{fig:4}
\end{figure}

\begin{figure}
\centerline{\epsfig{file=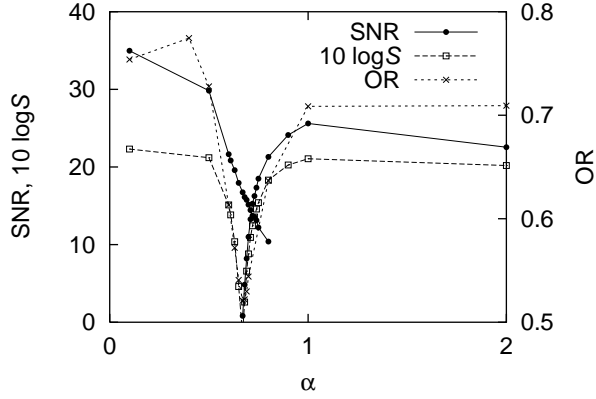,width=8cm}}
\caption{SNR, $10\log S$ (left vertical scale), and OR (right scale) 
at $T_{SR}$ as functions of the anisotropy factor $\alpha$.}
\label{fig:5}
\end{figure}

\begin{figure}
\centerline{\epsfig{file=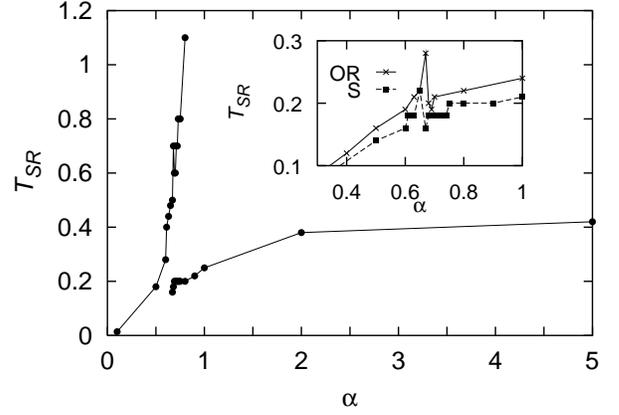,width=8cm}}
\caption{Resonance temperature $T_{SR}$, obtained from the SNR, 
as a function of the anisotropy factor $\alpha$.
It is ill-defined near $\alpha \simeq \alpha_{cr}$, where
the system response is strongly suppressed.
The two values of $T_{SR}$ for $\alpha \gtrsim \alpha_{cr}$
correspond to the double-peak structure of the SNR (see Fig.~\ref{fig:7}).
Behavior of $T_{SR}$ from the occupancy ratio as well as from the output
signal strength is shown in the inset.}
\label{fig:6}
\end{figure}

We next turn to the anisotropy effects, which are summarized in
Figs.~\ref{fig:4}, \ref{fig:5}, and \ref{fig:6}:
(i) The OR undergoes a dramatic change as $\alpha$ is
varied over $\alpha_{cr} \simeq 0.67$.
For $\alpha < \alpha_{cr}$ the OR reaches its minimum values 
at the resonance temperature $T_{SR}$ whereas maximum values are
attained for $\alpha > \alpha_{cr}$ (Fig.~\ref{fig:4}).
The OR minimum corresponds to a phase difference of approximately $\pi$
between the applied field and the staggered magnetization.
(ii) The SNR at $T_{SR}$ is substantially suppressed near $\alpha \approx
\alpha_{cr}$, as shown in Fig.~\ref{fig:5}.
(iii) The resonance temperature $T_{SR}$ in general increases with $\alpha$, 
although it is ill-defined around $\alpha_{cr}$ 
(Fig.~\ref{fig:6}; see also below).
These curious behaviors can be understood by recalling that the
anisotropy factor affects the system in two different ways:
On one hand, $\alpha$ simulates the effective coupling strength between
plaquettes [see Eq.~(\ref{Gx})]; this is analogous to the coupling strength
in the system of bistable oscillators, and explains straightforwardly 
the increase of $T_{SR}$ with $\alpha$.
On the other hand, depending on the value of $\alpha$, the system favors
one of the two states $\Psi_\pm$ [see the discussion below Eq.~(\ref{df})],
giving rise to the $\pi$-shift-like behavior in the system 
response (occupancy ratio) with respect to the signal.
In addition, for $\alpha \simeq \alpha_{cr}$, the external currents are not
efficient to drive the system since it is not obvious which links the
currents will prefer. This naturally leads to suppression of the
system response, as shown in Fig.~\ref{fig:4} and more clearly in
Fig.~\ref{fig:5}.
Note the rather reasonable agreement between 
the crossover anisotropy $\alpha_{cr}$ in simulations and
$\alpha_{cr}^0=1/2$ obtained from simple consideration 
of Eq.~(\ref{df}).

\begin{figure}
\centerline{\epsfig{file=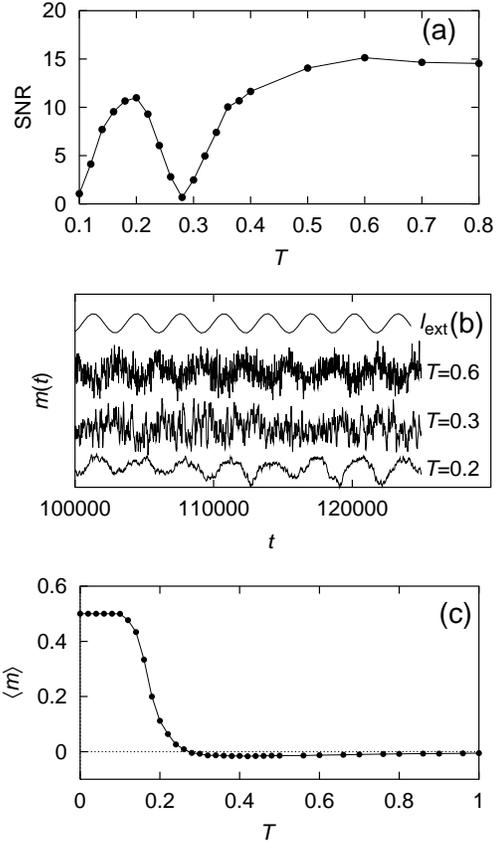,width=8cm}}
\caption{ Reentrance-like behavior near $\alpha_{cr}$: 
(a) the SNR as a function of the temperature for $\alpha = 0.7$
and (b) the corresponding time-series of the staggered magnetization $m(t)$.
For clarity, vertical positions of the data curves are shifted with 
respect to each other.
In (c), also shown is the equilibrium value of the staggered magnetization as a
function of the temperature in the presence of static staggered currents
($\Omega=0$).}
\label{fig:7}
\end{figure}

We now concentrate on the response of the system around $\alpha \approx
\alpha_{cr}$.  Surprisingly, Fig.~\ref{fig:7}(a) reveals that the SNR for
$\alpha=0.7$, slightly above $\alpha_{cr}$, exhibits reentrance-like
behavior, characterized by a double peak.  Further, the time series of the
staggered magnetization shown in Fig.~\ref{fig:7}(b) indicate that the
system favors different states at lower and at higher temperatures.
One possible explanation for these effects goes as follows: For
$\alpha\gtrsim\alpha_{cr}$, the currents at a given time $t$ favor to flow
through legs, but only slightly, with $\Psi_\pm$ almost degenerate. 
In our model described by Eqs.~(\ref{motion}) and (\ref{noise}), 
the ratio of the noise
level to the Josephson coupling on each link is given by $T$ on rungs and
$T/\alpha$ on legs, resulting in that the legs are subject to stronger random
noise than the rungs ($\alpha_{cr}\lesssim\alpha<1$).  As the temperature
is raised, stronger random noises thus effectively hinder the currents from
flowing through the legs leading to opposite circulation currents on
neighboring plaquettes.  This argument is consistent with the equilibrium
value of the staggered magnetization $\langle{m}\rangle$ as a function of
the temperature in the presence of a {\em static} staggered current
($\Omega=0$), displayed in Fig.~\ref{fig:7}(c).
For those values of $\alpha$, which lead to the double-peak structure, both
temperatures corresponding to the two peaks have thus been plotted in
Fig.~\ref{fig:6}.

\section{Analytical Results}

In this section we examine the SR behavior in an analytical way,
based on the Glauber dynamics~\cite{Glauber} of vortices.
Since the effective interaction between vortices, given by Eq.~(\ref{Gx}),
decays exponentially with the interaction range of the order of the 
lattice constant, we can regard it essentially as the nearest-neighbor interaction.
The effective vortex Hamiltonian in Eq.~(\ref{VHam2}), together with
Eq.~(\ref{tdep}), then reduces to
the one-dimensional (ferromagnetic) Ising Hamiltonian in the oscillating field
\begin{equation} \label{HI}
H_I = - J_I \sum_x \sigma_x \sigma_{x+1} - h \sum_x \sigma_x
\end{equation}
with the (staggered) Ising spin $\sigma_x \equiv 2(-1)^x m_x$, where
the nearest-neighbor interaction and the oscillating field are given by
\begin{eqnarray}
J_I &\equiv & {\pi^2  \over 2}
{ 1 + \alpha - \sqrt{1+2\alpha} \over \sqrt{1+2\alpha} }, \nonumber \\
h &\equiv& - 2 \pi { (2\alpha -1)(1+\alpha - \sqrt{1+2\alpha} ) 
\over (1+4\alpha^2)\sqrt{1+2\alpha}  } I_0 \cos\Omega t.
\end{eqnarray}

We consider the Glauber dynamics of the Hamiltonian (\ref{HI})
and compute the SNR of the staggered magnetization
$m(t)=(1/2L)\sum_x \sigma_x$.
We follow the procedure in Ref.~\onlinecite{Siewert} and 
obtain the output signal strength and the noise level
\begin{eqnarray} \label{SN}
S &=& {\pi h^2 \over 8T^2 } {A^2 (1-\gamma^2) \over A^2 (1-\gamma)^2 + \Omega^2 },
 \nonumber \\
N(\Omega) &=& {1\over2} 
\Re \left[ {1 + \eta s_2 \over s_1 (1-\eta s_2 ) }\right],
\end{eqnarray}
where $A\equiv\exp(-J_I/T)$, $\gamma \equiv \tanh (2J_I / T)$,
$\eta\equiv\tanh (J_I/T)$, $s_1 \equiv [(A+ i \Omega)^2 - A^2 \gamma^2]^{1/2}$,
$s_2 \equiv A \gamma / (A+i\Omega + s_1 )$,
and $\Re$ denotes the real part.
 
\begin{figure}
\centerline{\epsfig{file=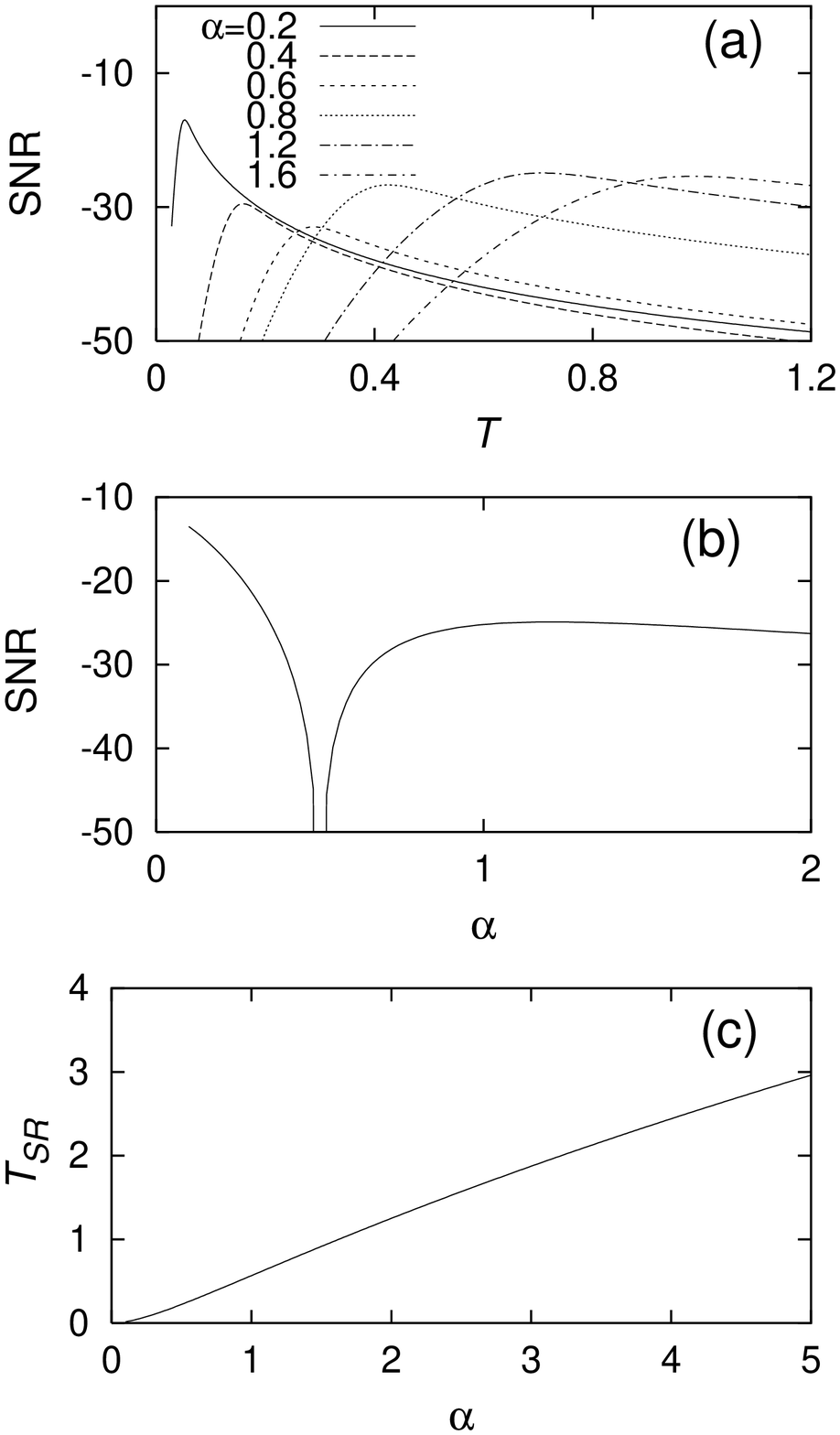,width=8cm}}
\caption{Resonance behavior displayed by the Glauber dynamics of vortices 
in FFJJL:
(a) SNR as functions of temperature for various values of $\alpha$;
(b) SNR at $T_{SR}$ as a function of the anisotropy factor $\alpha$;
(c) $T_{SR}$ as a function of the anisotropy factor $\alpha$.}
\label{fig:8}
\end{figure}

The SNR given by Eq.~(\ref{SNR}) together with Eq.~(\ref{SN}) 
is plotted in Fig.~\ref{fig:8}(a) 
for various values of $\alpha$, demonstrating clearly
the SR behavior.
Furthermore, it reproduces most of the anisotropy effects, which have been
observed in the numerical simulations. 
For example, Fig.~\ref{fig:8}(b) reveals substantial suppression of the SNR 
at $T_{SR}$ for $\alpha \approx \alpha_{cr}$, while the increasing tendency 
of $T_{SR}$ with $\alpha$ is shown in Fig.~\ref{fig:8}(c).
Interestingly, however,
in spite of the qualitatively good agreement described above,
one striking difference can be recognized between the analytical
results based on the Glauber dynamics and the simulation results 
obtained from the RSJ (Langevin) dynamics:
In the former there is no reentrance-like behavior in the SNR and
consequently no abrupt increase of $T_{SR}$ near $\alpha \approx \alpha_{cr}$.
Further, the crossover anisotropy close to the value
$\alpha_{cr}^0=1/2$ is obtained, while the simulations 
of the RSJ dynamics have given $\alpha_{cr}\approx{}0.67$.

\section{Conclusion}
We have considered a fully frustrated Josephson-junction ladder, driven by
external currents, periodic in time and staggered in space.  In particular
the spatiotemporal aspects of the stochastic resonance, resulting from the
interplay between the external driving and the thermal noise, have been
examined by means of numerical simulations performed on the resistively
shunted dynamics.  It has been found that the anisotropy in the Josephson
coupling along the leg- and the rung-direction yields rich physics,
e.g., resulting in a double peak SR structure.  We have also presented
an analytical investigation, based on the Glauber dynamics of vortices, and
which is in a qualitative agreement with with the simulation results. 

A particularly interesting aspect of the fully frustrated Josephson junction
ladder is the possibility of manufacturing such a structure and performing
real measurements.
Owing to the recent rapid advancement in lithographic technology, already
available are large-sized Josephson-junction ladders of good quality.  
In order to verify the SR effects discussed in the present paper
one might measure vortices  on (possibly selected)
plaquettes in the ladder using the superconducting quantum interference
devices (SQUID) with a wide range of operational frequency.~\cite{Shaw}
Another indirect way might be to measure the voltages across as many
selected legs as possible.
Highly challenging from an experimental point of view would be
to apply a staggered (but otherwise uniform) bias
current (through each site) along the array.\cite{Majer} It requires an
impedance for each injection of current, which should be sufficiently large
compared with the normal-state junction resistance.  One needs to choose
superconducting materials with higher Josephson energy, which allows higher
operation temperatures and hence more choices for resistors.  Another
possibility could be to use an auxiliary array of junctions, which provides
high impedances at sufficiently high frequencies.\cite{Majer}

\acknowledgments

MSC acknowledges support from Ume\aa\ University during his visit, where
main part of this work was done, and thanks J.\ Majer for
valuable discussions about possible experiments and C.\ Bruder for
useful discussions.  This work was supported in part by the Swedish Natural
Research Council through contract FU 04040-332 (BJK, MSC, PM) and by the
Ministry of Education of Korea through the BK21 Program (HJK, MYC).

\appendix
\section{}

In this Appendix, we derive the effective vortex Hamiltonian (\ref{VHam2}).
We start with the static case and consider for
convenience the $XY$ Hamiltonian:
\begin{equation} \label{Ham1}
H = - \sum_{\langle i,j \rangle} J_{ij} \cos (\phi_{ij} - A_{ij}),
\end{equation}
which corresponds to Eq.~(\ref{motion}) in the absence of external current
driving ($I_0 = 0$).  Application of the duality
transformation~\cite{skim,Jose} at sufficiently low temperatures then
decomposes the Hamiltonian in Eq.~(\ref{Ham1}) into the sum $H = H_{SW} +
H_V$ of the spin-wave part $H_{SW}$ (which is given in a trivial form) and
the vortex part
\begin{equation} \label{VHam}
H_V = \sum_{x,x'} ( v_x - f )G(x-x')(v_{x'} - f ).
\end{equation}

In the presence of static staggered currents [ $I_0 \ne 0$ but $\Omega=0$ 
in Eq.~(\ref{current})], we adopt the Villain scheme~\cite{Villain} to
obtain, to the linear order in $I_0$, the effective vortex Hamiltonian
\begin{equation} \label{VHam3}
H_V = \sum_{x,x'} ( v_x - f - \delta f_x)G(x-x')(v_{x'} - f - \delta
f_{x'}), 
\end{equation}
where the additional effective flux $\delta f_x$ depends on
the fraction $\rho$ of the external currents flowing
through rungs in the zero-temperature configuration:
\begin{equation}
\delta f_x = {1 \over \pi} \left( {1 -\rho \over 2 \alpha} - \rho
\right) (-1)^x I_0.
\end{equation}
>From Eq.~(\ref{motion}) without noise current, 
$\rho$ can be obtained
\begin{equation}
\rho = {1 \over 1 + 4 \alpha^2 },
\end{equation}
leading to
\begin{equation} \label{df}
\delta f_x = {  2 \alpha -1 \over \pi ( 1 + 4 \alpha^2)} (-1)^x I_0.
\end{equation}

When the external (staggered) current is periodic in time $(\Omega \ne 0)$, 
the additional flux also varies periodically in time. 
For sufficiently low-frequency driving ($\Omega \ll 1$), 
we can take Eq.~(\ref{tdep}) for the additional time-dependent flux $\delta
f_x(t)$.
Such a periodic flux induces oscillations between the two states
$\Psi_\pm$, suggesting the possibility of the SR behavior in the system.

%

%

\begin{references}

\bibitem[*]{gun} Present address: Center for Strongly Correlated Materials
 Research, Seoul National University, Seoul 151-742, Korea.
\bibitem{Benzi}
R. Benzi, A. Sutera, and A. Vulpiani, J. Phys. A {\bf 14}, L453 (1981).
\bibitem{Gammaitoni}
L. Gammaitoni, P. H\"anggi, P. Jung, and F. Marchesoni, Rev. Mod. Phys.
{\bf 70}, 223 (1998).
\bibitem{Lindner}
J.F. Lindner, B.K. Meadows, W.L. Ditto, M.E. Inchiosa, and A.R. Bulsara,
Phys. Rev. Lett. {\bf 75}, 3 (1995);
J.F. Lindner, B.K. Meadows, W.L. Ditto, M.E. Inchiosa, and A.R. Bulsara,
Phys. Rev. E {\bf 53}, 2081 (1996);
F. Marchesoni, L. Gammaitoni, and A.R. Bulsara, Phys. Rev. Lett. {\bf 76},
2609 (1996).
\bibitem{Locher}
M. L\"ocher, G.A. Johnson, and E.R. Hunt, Phys. Rev. Lett. {\bf 77}, 4698
(1996).
\bibitem{Benzi2}
R. Benzi, A. Sutera, and A. Vulpiani, J. Phys. A {\bf 18}, 2239 (1985).
\bibitem{book}
{\it Nonequilibrium Statistical Mechanics in One Dimension}, edited by V.
Privman (Cambridge University Press, Cambridge, 1997).
\bibitem{Neda}
Z. N\'eda, Phys. Lett. A {\bf 210}, 125 (1995);
Z. N\'eda, Phys. Rev. E {\bf 51}, 5315 (1995);
S.W. Sides, P.A. Rikvold, and M.A. Novotny, {\it ibid}. {\bf 57}, 6512
(1998);
K.-T. Leung and Z. N\'eda, {\it ibid}. {\bf 59}, 2730 (1999).
\bibitem{Tinkham}
M. Tinkham, {\it Introduction to Superconductivity}, 2nd ed. (McGraw-Hill,
New York, 1996).
\bibitem{Benz}
S.P. Benz, M.S. Rzchowski, M. Tinkham, and C.J. Lobb, Phys. Rev. Lett. {\bf
64}, 693 (1990).
\bibitem{Granato}
E. Granato, Phys. Rev. B {\bf 42}, 4797 (1990).
\bibitem{Denniston}
C. Denniston and C. Tang, Phys. Rev. Lett. {\bf 75}, 3930 (1995).
\bibitem{Kim} 
B.J. Kim, S. Kim, and S.J. Lee, Phys. Rev. B {\bf 51}, 8462 (1995); 
W.G. Choe and S. Kim, {\it ibid}. {\bf 53}, R502 (1996).
%
\bibitem{endnote}
In common experimental situation, the anisotropy in the Josephson
energy $J$ determines that in the normal-state resistance $R$
since the BCS theory predicts
$J\propto{}(\Delta/R)\tanh(\Delta/k_B T)$, where $\Delta$ is the
superconducting gap energy. 
In more sophisticated setups, e.g., replacing
rung- or leg-junctions by dc-SQUIDS, however, one can control the effective
Josephson energy independently of the normal-state resistance.
Thus, for simplicity, we here
assume the same normal-state resistance for rungs and legs.
%
\bibitem{Press} 
W.H. Press, S.A. Teukolsky, W.T. Vetterling, and B.P.  Flannery, 
{\it Numerical Recipes in C: The Art of Scientific Computing},
2nd ed. (Cambridge University Press, Cambridge, 1992).
\bibitem{Glauber}
R.J. Glauber, J. Math. Phys. {\bf 4}, 294 (1963).
\bibitem{Siewert}
U. Siewert and L. Schimansky-Geier, Phys. Rev. E {\bf 58}, 2843 (1998).
\bibitem{Shaw}
T.J. Shaw, M.J. Ferrari, L.L. Sohn, D.-H. Lee, M. Tinkham, and J. Clarke,
Phys. Rev. Lett. {\bf 76}, 2551 (1996).

\bibitem{Majer} J. Majer, Private communication.

\bibitem{skim}
S. Kim, Phys. Lett. A {\bf 229}, 190 (1997).
\bibitem{Jose}
J.V. Jos\'e, L.P. Kadanoff, S. Kirkpatrick, and D.R. Nelson, Phys. Rev. B
{\bf 16}, 1217 (1977).
\bibitem{Villain}
M.Y. Choi and S. Kim, Phys. Rev. B {\bf 44}, 10411 (1991);
J. Villain, J. Physique {\bf 36}, 581 (1975).
\end{references}
\end{document}